# A MAC Layer Based Defense Architecture for Reduction-of-Quality (RoQ) Attacks in Wireless LAN

Jatinder Singh,

*Director, Universal Institute of*

*Engg. & Tech. Lalru-CHD(India)*

Dr. Savita Gupta,

*Prof. Deptt. Of Computer Engg.*

*UIET, PunjabUniversity, CHD (India)*

Dr. Lakhwinder Kaur,

*Reader, UCOE, Punjabi)*

*University,Patiala (India)*

## Abstract

*Recently an alternative of DDoS attacks called shrew attacks or Reduction-of-Quality (RoQ) has been identified which is very much difficult to detect. The RoQ attacks can use source and destination IP address spoofing, and they do not have distinct periodicity, and may not filter the attack packets precisely. In this paper, we propose to design the MAC layer based defense architecture for RoQ attacks in Wireless LAN which includes the detection and response stages. The attackers are detected by checking the RTS/CTS packets from the MAC layer and the corresponding attack flows are blocked or rejected. By our simulation results, we show that our proposed technique achieves reduces the attack throughput there by increasing the received bandwidth and reducing the packet loss of legitimate users.*

## 1. Introduction

The characteristic vulnerabilities of LAN (802.11x) networks makes the attack possible. Like wired networks, it is not possible to physically protect the wireless networks. The next offices, the parking lot of the building, across the street or possibly several miles away, are the source of attacks for the wireless networks, i.e. the attack can be carried out from anyplace. It is necessary to realize the facts of different attacks in opposition to the wireless infrastructure to establish the appropriate defense strategy. The execution of several attacks which are not dangerous can be performed easily while the attacks which have destructive effects can be seriously complicated. The risk which is included in the wireless security is more when compared with any other case of security.

There are many potentially disturbing threats to wireless local area networks (WLANs). The security issues which are ranging from misconfigured wireless access points (WAPs) to session hijacking to Denial of Service (DoS) affects the WLAN. Wireless networks are also vulnerable on the particular threats in the wide array of 802.11, other than the TCP/IP based attacks which are related to the wired networks. A security solution which includes an intrusion detection system (IDS) should be used by the WLANs in order to assist in the defense and detection of these possible threats. Even organizations not having a WLAN should consider an IDS solution since it may be dangerous due to wireless threats [2].

WLAN are exposed to a variety of threats. The standard 802.11 encryption method, known as Wired Equivalent Privacy (WEP), is weaker.

The hackers attack a WLAN and collect the sensitive data by introducing a misbehaving WAP into the WLAN coverage area. The misbehaving WAP can be designed like an actual WAP because several wireless clients are connected simply to the WAP with the best signal strength. Moreover, the users can be "trapped" to connect with the misbehaving WAP unintentionally. When a user is associated, all the communications can be monitored by the hacker through the misbehaving WAP. Apart form the hackers, misbehaving nodes can also be introduced by the users. When low cost and easy implementation is combined with the flexibility of the wireless network communications, the WLANs can be very attractive to the users. By installing a WAP on a recognized LAN, the users can create a backdoor into the network, undermining all the hardwired security solutions and give the network open to the hackers [2].

Since the networks which are using 802.11 are vulnerable to numerous Denial of service (DoS) attacks, WLAN can be made fatal. Wireless communications which are uncertain upon physical objects are naturally vulnerable to signal degradation. Also, the hackers can overflow WAPs with association requests and injects the malicious DoS attacks there by forcing them to reboot. Moreover, by sending a repeated disassociate/deauthenticate requests with the help of the above mentioned rouge WAP, the hackers can refuse service to a wireless client.

Still there are various threats to WLAN and the identification of further vulnerabilities is performed at a high speed. The general truths are the reality of the threats, their ability to create broad destruction and their rising familiarity with increase in fame of the 802.11 technology. With the absence of the detection mechanism, the identification of the threats to a WLAN can be complicated. When the consciousness of the threats is absent then a network is inadequately secure with respect to the threats facing it. When the threats to the networks are recognized, then it is possible to equip the WLAN suitably with the necessary security measures.

**1.1 RoQ Attack**

By the high rate or high volume, the typical DDoS flooding attacks are characterized. An alternative of DDoS attacks has been identified recently which is too complex to detect which are called as shrew attacks or Reduction-of-Quality (RoQ) attacks. Instead of refusing the clients from the services completely, these RoQ attacks throttle the TCP throughput heavily and reduce the QoS to end systems gradually [3].

The transients of systems adaptive behavior is targeted by the RoQ attacks instead of limiting its steady-state capacity. The RoQ attacks can use source and destination IP address spoofing, and they do not have distinct periodicity, and may not filter the attack packets precisely. In order to escape from being caught by the traceback techniques, RoQ attacks often launch attacks through multiple zombies and spoof header packet information. But, it is important to control the frequency domain characteristics of attacking flows. In order to throttle the TCP flows efficiently, the attacking period has to be close to the Retransmission Time Out (RTO). Using traffic spectrum, the energy distribution pattern will give up such malicious flow detection mechanisms even if the source IP addresses are carried in packet header are falsified [3].

In this paper, we propose to design a MAC layer based defense architecture for Reduction-of-Quality (RoQ) attacks in Wireless LAN. It includes the detection and isolation of attackers. Detection makes use of three status values that can be obtained from the MAC layer: Frequency of receiving RTS/CTS packets, frequency of sensing a busy channel and the number of RTS/DATA retransmissions. Once the attackers are detected, the corresponding flows are blocked or rejected.

**2. Related Work**

Yu Chen et al [3] have explored the energy distributions of Internet traffic flows in frequency domain. Normal TCP traffic flow present periodicity because of protocol behavior. Their results revealed that normal TCP flows can be segregated from malicious flows according to energy distribution properties. They have discovered the spectral shifting of attack flows from that of normal flows. Combining flow-level spectral analysis with sequential hypothesis testing, they have proposed a novel defense scheme against RoQ attacks. Their detection and filtering scheme can effectively rescue 99% legitimate TCP flows under the RoQ attacks.

Zhongua Zhang et al [4] focused on the examination of the anomaly-based intrusion detector's operational capabilities and drawbacks through their operating environments. Anomaly detection is classified in a statistical framework based on the similarity with the induction problem for describing their general expected behaviors. For the apparent subjects from hosts and networks, several key problems and respective potential solutions about the normality characterization for the observable subjects are addressed. Based on some existing achievements anomaly detector's evaluations are also examined.

Piyush Kumar Shukla et al [5] have analyzed the congestion based DDoS attacking in mobile ad hoc networks. They have proposed a grammar based approach to modeling and analyzing multi-step network attack sequences. Moreover, they have proposed a low rate DoS attack detection algorithm, which relies on the core characteristic of the low rate DoS attack in introducing high rate traffic for short periods, and then uses a proactive test based differentiation technique to filter the attack packets. They have evaluated the feasibility of the proposed low rate DoS attack algorithm on real Internet traces.

Martin Rehak et al [6] have proposed a research to detect malicious traffic in high-speed networks by correlated anomaly detection methods. Based on FPGA elements transparent inline probes are used to obtain the real time traffic statistics in NetFlow format and gives a traffic statistics to the agent-based detection layer. The agent uses a particular anomaly detection method in this layer to detect the anomalies and describes the flows in its extended trust model. The agent shares the anomaly estimation of the individual network flows which are uses as an input for the agents trusts models. In order to estimate their maliciousness the trustfulness values of individual flows from all agents are combined.

John Haggerty et al [7] have proposed that a major threat to the information economy is denial-of-service attacks. These attacks are common even though the widespread usage of the perimeter model countermeasures. Therefore to provide early detection of flooding denial-of-service attacks, a new approach is assumed which uses statistical signatures at the router. There are three advantages for this approach. They are: Computational load on the defense mechanism is reduced by analyzing fewer packets, if the system is under protection then the state information is not required and alerts may span many attack packets. Thus to prevent malicious packets from reaching their proposed target in the first the defense mechanism may be placed within the routing infrastructure.

Mina Guirguis et al [8] have analytically captured the effect of RoQ attacks that would deprive an Internet element from reaching steady state by knocking it off whenever it is about to stabilize. They have formalized the notion of attack "potency", which exposes the tradeoff between the "damage" inflicted by an attacker and the "cost" of the attack. Moreover, their notion takes aggressiveness into account which enabled them to identify different families of DoS attacks based on their aggressiveness.

Mina Guirguis et al [9] have exemplified the security implications by exposing the vulnerabilities of admission control mechanisms that are widely deployed in Internet end systems to Reduction of Quality (RoQ) attacks. They have shown that a well orchestrated RoQ attack on an end-system admission control policy could introduce significant inefficiencies that could potentially deprive an Internet end-system from much of its capacity, or significantly reduce its service quality, while evading detection by consuming an unsuspicious, small fraction of that system's hijacked capacity. They have developed a control theoretic model for assessing the impact of RoQ attacks on an end-system's admission controller. They quantified the damage inflicted by an attacker through deriving appropriate metrics.

Eduardo Mosqueira-Rey et al [10] have described the design of misuse detection agent which is one of the different agents in a multiagent-based intrusion detection system. Using a packet sniffer the agent examines the packets in the network connections and creates a data model based on the information obtained. This data model is the input to the rule based agent inference engine which uses the Rete algorithm for pattern matching. So the rules of the signature-based intrusion detection system become small.

Magnus Almgren et al [11] have investigated the procedure to use the alerts from may audit sources to improve the accuracy of the intrusion detection system (IDS). A theoretical model is designed automatically for the reason about the alerts from the different sensors through concentrating on the web server attacks. It also provides a better understanding of possible attacks against their systems for the security operators. This model enables reasoning about the absence of the expected alerts by taking the sensor status and its capability into account. This model is built using Bayesian networks which needs some initial parameter values that can be provided from the IDS operator.

Naeimeh Laleh et al [12] have proposed that fraud is growing remarkably with the growth of modern technology and the universal superhighways of communication which results in the loss of billions of dollars throughout the world each year. This technique tends to propose a new taxonomy and complete review for the different types of fraud and data mining techniques of fraud detection. The uniqueness of this technique is gathering all types of frauds which can be detected by data mining techniques and analyzes some real time approaches which have the ability to detect the frauds in real time.

Yu Chen et al [13] have proposed a new signal-processing approach to identify and detect the attacks by examining the frequency domain characteristics of incoming traffic flows to a server. Their proposed technique is effective in that its detection time is less than a few seconds. Furthermore, their technique entails simple implementation, making it deployable in real-life network environments.

## 3. Proposed Defense Technique

In this paper, we propose a defense scheme that includes the detection and response stages.

Detection makes use of three status values that can be obtained from the MAC layer: frequency of receiving RTS/CTS packets, frequency of sensing a busy channel, and the number of RTS/DATA retransmissions. When the number of RTS/CTS packets received exceeds a certain threshold $RC_{th}$, it indicates that too many nodes are within the transmission range to compete for the channel. When the channel is sensed to be in a busy state, a node will persist in the backoff stage and stop the CW count. When the stopping time is longer than a threshold $SE_{th}$, it indicates that too many nodes are within the interference range. Thus if the number of retransmissions exceeds a threshold $RE_{th}$, it will be regarded as an indicator for channel congestion. Since these status values are already available in the protocol stack implementation, the overhead required for implementing this detection scheme is very low.

During the response phase, the nodes will check the following conditions to mark each packet with a Congestion Bit (CB)

1. If number of RTS/CTS packets > $RC_{th}$,
2. If Stime > $SE_{th}$,
3. If number of RTS/DATA retransmissions > $RE_{th}$,

Now, the congestion bit is set as follows:

CB = 000: If none of the Conditions are true.

CB = 100: If Condition 1 is true.

CB = 010: If Condition 2 is true.

CB = 001: If Condition 3 is true.

CB = 110: If Conditions 1&2 are true.

CB = 101: If Conditions1&3 are true.

CB = 011: If Conditions 2&3 are true.

CB = 111: If all the Conditions are true.

We propose that RTS and CTS values between two communicating nodes can be observed by a passive server and it calculates the CB. This helps to detect and isolate the attackers.

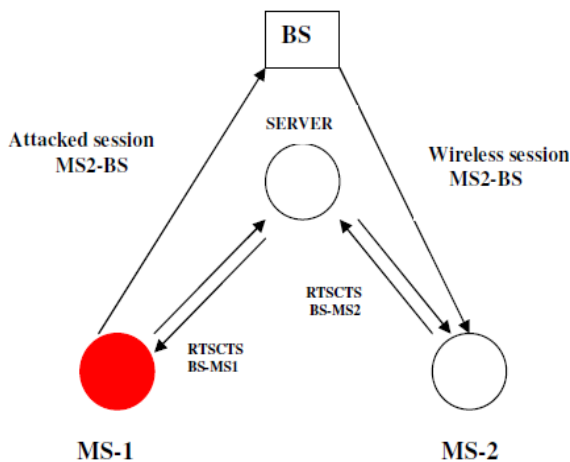

Fig. 1 RTS/CTS monitoring by Server

The Server executes the following algorithm, to detect the attackers.

**Algorithm**

1. Let initial time interval = t1

2. Server checks RTS/CTS packets and calculates CB.

3. If CB ≠ 000, then

   3.1 If CB = 100 or 010 or 001, then

      3.1.1 Mark the source status as normal and transmit CB to the source.

   3.2 Else if CB = 110 or 101 or 011, then

      3.2.1 Mark the source status as suspected and transmit CB to the source.

   3.3 Else if CB = 111, then

      3.3.1 Mark the source status as attacker and transmit CB to the source.

   3.4 End if

4. End if

5. t1 = t1 + 1

6. Repeat the steps from 2

7. If t1 = 3 and source status = attacker, then

   6.1 Remove the node from the list.

   6.2 Block all the traffic from the attacker

8. Else If t1 = 4 and source status = suspected, then

   8.1 Remove the node from the list.

   8.2 Block all the traffic from the attacker

9. End if

At the initial time interval t1, the server checks the RTS/CTS packets and calculates the Congestion Bit (CB) according to the conditions mentioned above. If the value of CB is either 100 or 010 or 001, then the status of the source is marked as normal and the CB is transmitted to the source. If the value of CB is either 110 or 101 or 011, then the status of the source is marked as suspected and the CB is transmitted to the source. If the value of CB is 111 the status of the source is marked as attacker and transmits the CB to the source. Suppose, if the status of the source is still attacker till the time interval t3, then the corresponding node is removed from the list and all the traffics from the attackers are blocked. Similarly, suppose if the status of the source is still suspected till the time interval t4, then the corresponding node is removed from the list and all the traffics from the attackers are blocked.

**4. Simulation Results**

**4.1 Simulation Model and Parameters**

This section deals with the experimental performance evaluation of our algorithm through simulations. In order to test our protocol, the NS2 [14] simulator is used. We compare our proposed MAC Layer Based Defense Architecture for Reduction-of-Quality (RoQ) with Shrew [13] filter.

### 4.2 Performance Metrics

In our experiments, we measure the following metrics

- Received Bandwidth
- Packet Loss

The simulation results are described in the next section.

### 4.3 Results

#### A. Effect of Varying Attackers

In our first experiment, we vary the number of attackers as 2, 4, 6 and 8 in order to calculate the received bandwidth and packet loss legitimate users.

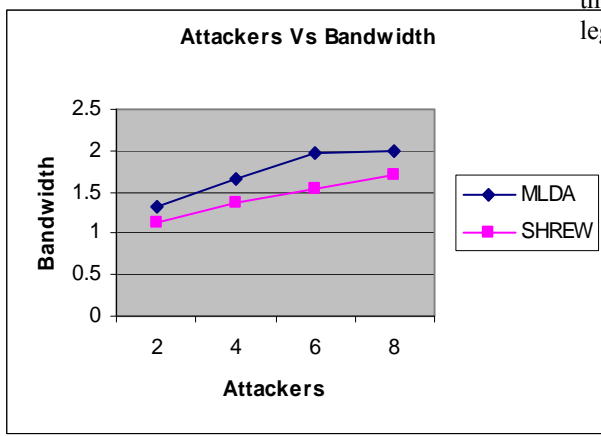

Fig: 2 Attackers Vs Bandwidth

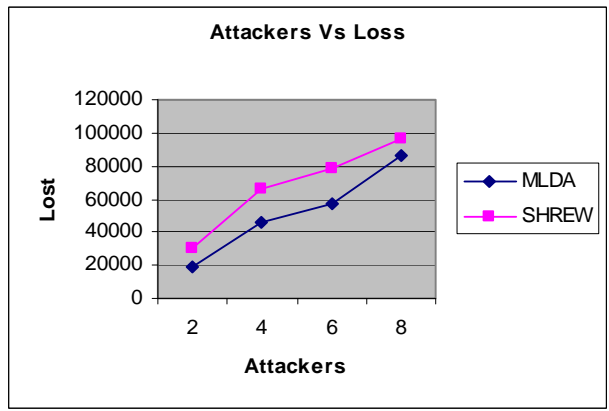

Fig: 3 Attackers Vs Loss

Fig: 2 gives the received bandwidth for normal legitimate users when varying the number of attackers. It shows that the bandwidth received for normal users is more in the case of MLDA when compared with SHREW.

Fig: 3 illustrates that the packet loss due to attack is more in SHREW when compared with MLDA, when varying the number of attackers.

.

#### B. Effect of Varying Attack Period

In our final experiment, we vary the number of attack period as 0, 5…20 in order to calculate the received bandwidth and packet loss of the legitimate users.

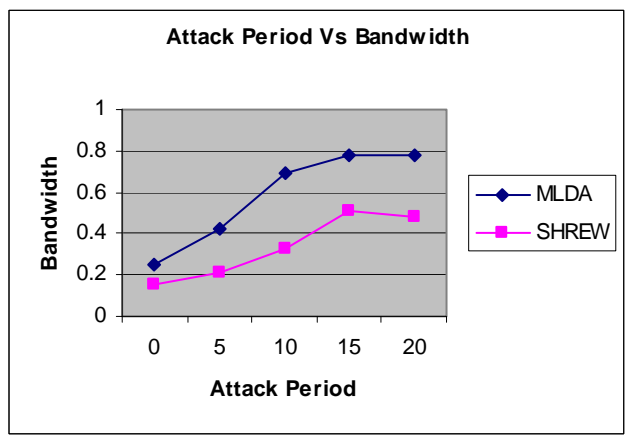

Fig: 4 Attack Period Vs Bandwidth

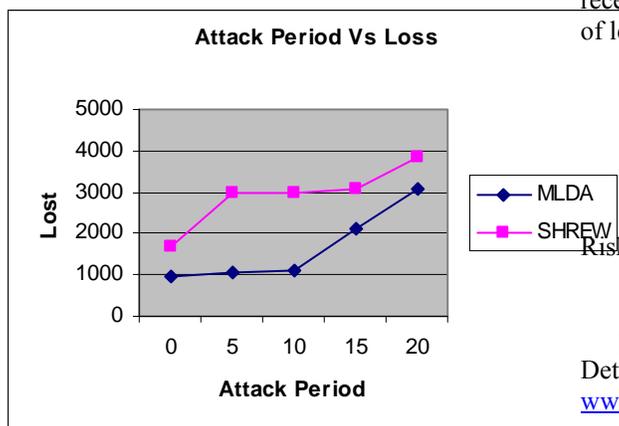

Fig: 5 Attack Period Vs Loss

Fig: 4 gives the received bandwidth for normal legitimate users when varying the number of attack period. It shows that the bandwidth received is more in the case of MLDA when compared with SHREW.

Fig: 5 illustrates that the packet loss due to attackers is more in SHREW when compared with MLDA.

## 5. Conclusion

In this paper, to defend against the Reduction-of-Quality (RoQ) attacks, we have proposed a MAC layer based defense architecture in Wireless LAN. It includes the detection and response stages. Detection makes use of three status values that can be obtained from the MAC layer: frequency of receiving RTS/CTS packets, frequency of sensing a busy channel, and the number of RTS/DATA retransmissions. In our proposed architecture, RTS and CTS values between two communicating nodes can be observed by a passive server to detect and isolate the attackers. It calculates a cumulative congestion bit (CB), depending on these three status values. By executing our algorithm the server detects the attackers. Once the attackers are detected, the corresponding attack flows are blocked or rejected. By our simulation results, we have shown that our proposed technique reduces the attack throughput there by increasing the received bandwidth and reducing the packet loss of legitimate users.

**References**

[1] Kimmo Hiltunen, "WLAN Attacks and Risks", White Paper, Ericson, January 2008.

[2] Jamil Farshchi, "Wireless Intrusion Detection System", Security Focus, Nov- 2003, www.securityfocus.com/infocus/1742

[3] Yu Chen and Kai Hwang, "Spectral Analysis of TCP Flows for Defense against Reduction-of-Quality Attacks", in the IEEE International Conference on Communications-ICC-2007, June 2007.

[4] Zonghua Zhang and Hong Shen, "A Brief Observation Centric Analysis on Anomaly Based Intrusion Detection", Springer-Verlag Berlin, Heidelberg 2005.

[5] Piyush Kumar Shukla, S. Silakari and S.S. Bhadouria, "Designing And Analysis Issues For An Attack Resilient and Adaptive Medium Access Control Protocol for Computer Networks: An Exclusive Survey",

[6] Martin Rehak, Michal pechoucek, karel Bartos, Martin Grill, Pavel celeda and vojtech krmick "An intrusion detection system for high-speed networks", national institute of informatics, 2008.

[7] John Haggerty, Qi Shi and Madjid Merabti, "Statistical Signatures For Early Detection Of Flooding Denial-Of-service Attacks", Springer Boston, 2006.

[8] Mina Guirguis, Azer Bestavros and Ibrahim Matta, "Exploiting the Transients of Adaptation for RoQ Attacks on Internet Resources**",** 12th IEEE International Conference on Network Protocols (ICNP'04).

[9] Mina Guirguis, Azer Bestavros, Ibrahim Matta and Yuting Zhang, "Reduction of Quality (RoQ) Attacks on Internet End-Systems", Proceedings IEEE 24th Annual Joint Conference of the IEEE Computer and Communications Societies, INFOCOM 2005.

[10] Eduardo Mosqueira-Rey, Amparo Alonso-Betanzos, Belen Baldonedo Del Rio, and Jesus Lago Pineiro, "A Misuse Detection Agent for Intrusion Detection in a Multi-agent Architecture", Springer-Verlag, Berlin Heidelberg 2007.

[11] Magnus Almgren, Ulf Lindqvist, and Erland Jonsson, "A Multi-Sensor Model to Improve Automated Attack Detection", Springer-Verlag Berlin Heidelberg 2008.

[12] Naeimeh Laleh and Mohammad Abdollahi Azgomi, "A Taxonomy of Frauds and Fraud Detection Techniques", Springer-Verlag Berlin Heidelberg 2009.

[13] Yu Chen, Yu-Kwong Kwok, and Kai Hwang, "Filtering Shrew DDoS Attacks Using A New Frequency-Domain Approach", In the IEEE Conference on Local Computer Networks, 2005, 30th Anniversary.

[14] Network Simulator: www.isi.edu/ns

## About the Authors

*Jatinder Singh* Received his M.Tech degree from Punjabi University, Patiala in 2003. He is a dynamic researcher and prolific author in the field of Computer Engineering. He has also won Best Research Scholal Award by UGC and Management Excellence Award by MIDI, Punjab. He published 50 National and International research papers over the years as well as 20 highly acclaimed text and research books. He is also a member of several professional scientific organizations and has lectured widely at academic institutions in India and Abroad. Presently is working as a Director in Universal Institute of Engg. & Tech. Lalru CHD.

*Lakhwinder Kaur* received the M.E. degree from TIET, Patiala, Punjab, in 2000 and Ph.D. degree from PTU, Jalandhar in 2007 both in computer science and engineering. She has been in the teaching profession since 1992. Presently, she is working as Reader in the Department of CSE, University College of Engg., Punjabi University, Patiala (Pb). Her research interests include image compression and denoising, Grid computing and wavelets.

*Savita Gupta* received the B.Tech. degree from TITS, Bhiwani (Haryana), in 1992, M.E. degree from TIET, Patiala, Punjab, in 1998 both in computer science and engineering. She obtained her Ph.D. degree from PTU, Jalandhar in 2007 in the field of Ultrasound Image Processing. She has been in the teaching profession since 1992. Presently, she is working as Professor in the Department of CSE, University Instiute of of Engg. & Technology, Panjab University, Chandigarh. Her research interests include image processing, image compression and denoising, and wavelet applications

--------------